\documentclass[conference]{IEEEtran}
\IEEEoverridecommandlockouts

\usepackage{cite}
\usepackage{amsmath,amssymb,amsfonts}
\usepackage{graphicx}
\usepackage{textcomp}
\usepackage{xcolor}
\def\BibTeX{{\rm B\kern-.05em{\sc i\kern-.025em b}\kern-.08em
    T\kern-.1667em\lower.7ex\hbox{E}\kern-.125emX}}

\usepackage{amsmath,amsfonts}
\usepackage{graphicx}
\usepackage{textcomp}
\usepackage{bmpsize}
\usepackage{xcolor}
\usepackage{lipsum}
\usepackage{algorithm} 
\usepackage{algpseudocode} 
\usepackage{subcaption}
\usepackage{graphicx}
\usepackage{booktabs}
\usepackage{mathrsfs}
\usepackage{multirow}
\usepackage{comment}
\usepackage{url}

\begin{document}
\title{Supporting Socially Constrained Private Communications with SecureWhispers}
\author{
\IEEEauthorblockN{Vinod Khandkar}
\IEEEauthorblockA{University of Surrey \\ v.khandkar@surrey.ac.uk
}
\and
\IEEEauthorblockN{Kieron Ivy Turk}
\IEEEauthorblockA{University of Surrey \\ k.turk@surrey.ac.uk
}
\and
\IEEEauthorblockN{Ehsan Toreini}
\IEEEauthorblockA{University of Surrey \\ e.toreini@surrey.ac.uk
}
\and
\IEEEauthorblockN{Nishanth Sastry}
\IEEEauthorblockA{University of Surrey \\ n.sastry@surrey.ac.uk
}
}
\maketitle
\begin{abstract}
Rapidly changing social norms and national, legal and political conditions socially constrain people who wish to discuss sensitive topics such as abortion, sexuality, or religion. Vulnerable minorities who are constrained or oppressed in this way are often worried about inadvertent information disclosure and may be unsure about the extent to which their communications are being monitored in public or semi-public spaces like workplaces or cafes. Personal devices extend trust to the digital domain, making it desirable to have strictly private communication between trusted devices. Currently, messaging services like WhatsApp provide alternative means for exchanging sensitive private information, while personal safety apps such as Noonlight enable private signaling. However, these rely on third-party mechanisms for secure and private communication. There are cases where such third-party mechanisms are not accessible, including lack of internet connectivity, insecure access, or companion device connections. In these cases, it is challenging to achieve communication that is strictly private between two devices instead of user accounts without any dependency on third-party infrastructure. The goal of this paper is to support private communications by setting up a shared secret between two or more devices without sending any data on the network. We develop a method to create a shared secret between phones by shaking them together. Each device extracts the shared randomness from the shake, then conditions the randomness to 7.798 bits per byte of key material. This paper proposes three different applications of this generated shared secret: message obfuscation, trust delegation, and encrypted beacons. We have implemented the message obfuscation on Android as an independent app that can be used for private communication with trusted contacts. We also present research on the usability, design considerations, and further integration of these tools in mainstream services.
\end{abstract}


\section{Introduction}
\label{sec:intro} 
As societal norms change, there may be groups of people who feel \textit{socially constrained} about some topics, i.e., they find it difficult to communicate openly with trusted others about certain sensitive issues in public and semi-public spaces such as work places. For instance, in the rapidly evolving landscape of USA politics, abortion rights are being rapidly withdrawn following the overruling of Roe v. Wade in 2022, forcing people considering abortion to discuss their decision in private. Furthermore, marginalised groups such as the LGBTQ+ community face increasing restrictions on their rights and need to communicate securely when faced with political and governmental actions against them. In other countries, religious minorities may be subject to discrimination or persecution. For instance, the underground church for the new-converts to Christianity is illegal in Iran, with serious consequences for its members. The Amnesty International report on Iran's human rights in 2023 explicitly referred to the house church and new-converts ``being subjected to arbitrary arrest and punishments such as imprisonment and internal exile''.

Another form of socially constrained communication is technology-facilitated abuse (TFA): the pervasive use of home and personal devices to perpetuate domestic abuse. In a stereotypical abusive household, it is common for one person to be responsible for setting up the household technical infrastructure including network infrastructure (such as the WiFi router) and monitoring-capable devices including doorbell cameras, security systems, and smart assistants. This creates a power asymmetry between technologically skilled user(s) and the other members of the household. For example, the person who has set up the WiFi router may then abuse this position to observe network traffic of their partner, or use CCTV/smart assistant recordings to listen to conversations that happened whilst they were not at home. This creates a powerful adversary in the situation of TFA, where the survivor may feel unable to leave a relationship and yet lacks a safe and secure channel to reach out to someone outside for help without alerting their partner.

The goal of this paper is to develop PETs primitives that can serve as the basis for secure and private messaging between members of a minority or vulnerable group with their trusted friends. We expect that the vulnerable users are likely to be in control of their personal devices (e.g., phones), but might not be sure of the extent to which their communications can be monitored or noticed, even if only inadvertently. This ranges from a disapproving family member seeing a notification pop up about abortion on the vulnerable user's phone to IT staff at a workplace monitoring communications for security reasons and inadvertently discovering a colleague's sexuality. 

A strawman solution could be to encrypt the message using a shared secret, but to use steganography and hide encrypted messages in other content, so that it does not appear suspicious to the adversary if they check the user's phone. However, creating a shared secret without the adversary also learning it can be challenging due to surveillance (or monitoring, even if for other reasons and not directly targeting the vulnerable individuals). End-to-End Encrypted channels such as WhatsApp or Signal may not be appropriate as they require a trusted server to bootstrap and verify the user handles or phone numbers at the time of channel setup \cite{10.1145/3424302.3425909}; and minority or privacy-conscious users may feel unsafe about trusting external sources\footnote{Some platforms like WhatsApp offer QR-code based verification when two users are in physical proximity (https://faq.whatsapp.com/820124435853543), but this may be susceptible to a variant of QRLfishing~\cite[\S3.2.2]{heartfield2018protection}, shoulder surfing or CCTV capture.}. 

\begin{figure}[t]
  \centering
  \includegraphics[scale=0.4]{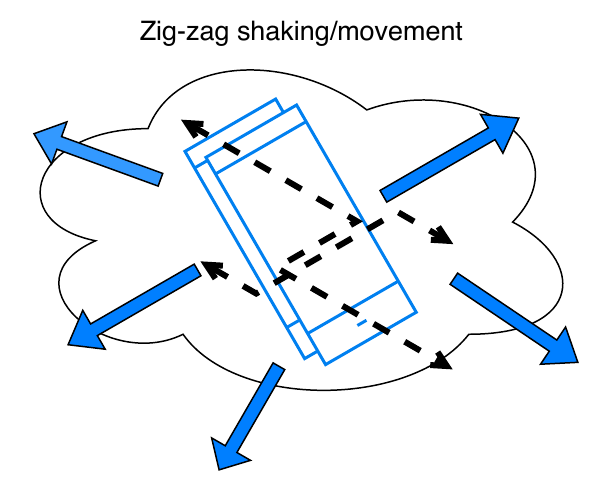}
  \caption{Core shared secret generation idea}
  \label{fig:basic_idea}
\end{figure}

Our key contribution is a method to generate shared secrets completely locally, by leveraging a common source of randomness shared by two devices. Our method requires no technical expertise from users: it only necessitates shaking two phones together (\textit{cf.} Fig.~\ref{fig:basic_idea}) and extracts shared randomness from the shared movements of phones to create a shared secret. 

The challenge for synchronising random movements across phones arises from the presence of noise in sensor readings. While the random gestures are supposed to generate similar sensor data on both phones, our experiments show the readings contain noise which lead to various discrepancies. We develop a Discrete Fourier Transform (DFT) based approach to filter out this noise and efficiently generate shared secrets. A cryptographic evaluation shows that the proposed method generates a shared entropy of over 6 bits per byte generated. This is comparable to OpenSSL's Real Time Clock-based entropy source for key generation, which generates 6.59–6.65 bits per byte \cite{8939840}. We further condition this entropy before using it as a key, achieving a total entropy of 7.798 bits per byte.

We experimentally validate that we can successfully generate shared keys by either aligning and holding the two phones together in one hand and shaking them (success rate of $92$--$92.6$\%), or by placing the phones together in a pocket and walking around (success rate between $86$--$91$\%). With a DFT size of 64 and $5$ input data variants we can generate more than 256 bits of entropy with one minute of physical movements of smartphones, which is considered to provide sufficient resilience against common brute-force attacks.  

As a proof-of-concept implementation, we develop ``SecureWhispers'', a tool that uses this method to set up secure channels of communication that are not only End-to-End Encrypted (like WhatsApp and other commercially available tools) but also the set up of the channel happens between two entities that are authenticated with each other. We co-designed SecureWhispers with a panel of eight people with lived experience of socially sensitive vulnerabilities and life transitions, who positively reviewed the final application for use in socially-constrained discussions.

The rest of the paper is organised as follows: Sec.~\ref{sec:rel_work} explains the previous works in related to shared secret generation and Sec.~\ref{sec:th_model} describes the threat model. Sec.~\ref{sec:chlngs} lists the challenges associated with generating secrets and Sec.~\ref{sec:method} describes the proposed secret generation. The evaluation results of the proposed secret generation method are present in Sec.~\ref{sec:eval}. Sec.~\ref{sec:securewhispersapp} describes the end-to-end SecureWhispers tool developed for generating a secret and secure messages, and 
the conclusion is in Sec.~\ref{sec:concl}.

\section{Related work}
\label{sec:rel_work}
Before identifying the challenges associated with smartphone-based secret generation, we present the background of sources of randomness and earlier work referring to smartphone-based secret generation. We additionally present related works on technology-facilitated abuse to develop our threat model in Sec.~\ref{sec:th_model}.

\subsection{Sources of randomness}
Creating a secure communication channel between two devices requires the shared secret agreed between two devices. One approach would be to use the certificates-based public-private \cite{sec_gen_rsa} key interface. However, the public-private-key method requires pre-setting specific secrets on both devices. Moreover, it results in network traffic associated with the secure channel establishment handshake, which can cause suspicion. 

Kouzani~\cite{ipv_tech_sol} lists various sources of randomness used in the existing technology solutions for domestic violence, such as data obtained from wearables, smartphones, and ambient parameters. The adversary's capability defines the suitability of the source of randomness. However, it underlines the challenges associated with using such technologies in a privacy-preserving manner for the privacy and safety of domestic abuse survivors.  

\subsection{Smartphone-based secret generation}
Various methods are developed in the existing literature to generate secrets over the non-covert communication channel. The communication channel dynamics are readily available for randomness in secret generation. Bhatti and Saleem~\cite{sec_gen_ephi} use the wireless channel dynamics observed between two devices as a source of randomness. Moreover, the method exhibits inherent snoop resilience as two pairs of devices observe channel dynamics differently. Safaka et al.~\cite{secret_oo_thinair} present a novel method to create and distribute secrets over wireless channels with induced noise on the channel instead of using existing channel dynamics. Both methods exchange necessary information over the wireless channel to distribute the randomness, which raises concerns about the safety of survivors. Our focus is on the smartphone-sensor-based secret generation method that distributes the randomness in a privacy-preserving manner. 

The sensor data is another source of randomness that can be synchronized between two devices to generate a shared secret. Leung et al.~\cite{10.1145/3214275}, Mare et al.~\cite{10.1145/3264935}, and Findling et al.~\cite{10.1145/2684103.2684122} present systems where two devices are shaken together and the smartphone-based sensors data measured. They compare the data for the similarity for authentication or unlocking one of the devices. However, these methods do not generate the shared secret or key. The work presented in this paper aims to generate a shared secret using smartphone sensor data. 

Other methods~\cite{10.1155/2021/6668478}, \cite{9417366}, \cite{10.1145/3378669}, \cite{mare:csaw19}, \cite{sec_gen_walkie_talkie}, \cite{sec_gen_fuzy_logic}, \cite{sec_gen_shake2com}, \cite{sec_gen_sens_pair}, \cite{10.1145/3023954} present secret generation based on the sensor data available on the smartphones or wearables. In these methods, the randomness generated by sensor data through similar movements of devices or capturing common events provides the foundation for the secret generation. While one approach extracts the shared randomness by event duration detection, another extracts the significantly random bits in sensor data. The final shared randomness or secret is agreed over the existing communication channels, such as Bluetooth. The key agreement over communication channels is undesirable from the survivor's safety and privacy perspective. The proposed method aims to generate the shared secret key without any communication between devices. 

The shared positional information-based secret generation approach is described by Vyas et al.~\cite{10211219}. Other ambient parameters use phase shifts due to artificial reflectors~\cite{9442833} and phase shifts at different antennas~\cite{10008695}. The secret agreement uses error correction followed by reconciliation over the existing communication channel. However, these methods also falls short on generating the shared secret without communication between devices.

The shared secret generation without any communication between smartphones is also a well researched topic. Bichler et al.~\cite{10.1007/978-3-540-74853-3_18,5227986} researched and successfully generated shared secret between two smartphones without any communication between smartphones. However, their method relies on the central computer to synchronise the acceleration data between smartphones. Their results demonstrate that the maximum entropy achieved with the proposed key generation algorithm is around 140 bits with single instance of generation of such key. However, this method succeeds in generating \textit{13 bits/key} with a success rate of $80\%$. The deployment scenario presented in our paper aims to generate a shared entropy with higher success rate and utilise it for user data encryption purpose. The encryption key requires higher entropy ($>$128 bits/key for symmetric key encryption/decryption~\cite{nist_800_131A}).  

\subsection{Technology-Facilitated Surveillance}\label{subsec:bg-tfa}
Technology-Facilitated Surveillance involves the use and abuse of technology for spying on other users. It manifests in many forms, from household surveillance to monitoring employees at work. One type of surveillance covered extensively in literature is Intimate partner surveillance (IPS), the most prevalent form of technology-facilitated domestic abuse. Tseng et al.~\cite{tseng20ipsinfidelityforums} identify a wide range of IPS methods from shared account access and router monitoring to recording devices through analysis of infidelity forums. Ceccio et al.~\cite{ceccio23spydevices} survey online retailers and find over 2\,000 different devices that can be used for spying on a partner, many of which explicitly advertise themselves for IPS. Stephenson et al.~\cite{stephenson23injail} focus on the abuse of IoT devices, noting a heavy focus on audio and video surveillance.

The key implications for our study are twofold. Firstly, adversaries in socially constrained communication scenarios often have access to shared devices and accounts (including their partner's mobile phone), which they can use to monitor a partner, and have methods to gain access to further devices for surveillance. Secondly, the most common form of abuse is IPS, involving a wide range of in-home and remotely accessible tools that allow for broader environment monitoring. This highlights the highly controlled environment that abuse survivors need to communicate out of without being detected despite all odds being against them. In our paper, we present a system which aids vulnerable people with similar communication challenges.

\section{Threat Model}
\label{sec:th_model}
SecureWhispers aims to address the needs of vulnerable people (we stereotype as `Alice') who may need to discuss sensitive topics but may feel socially constrained to, due to changing opinions, or emerging new political or legal norms. Examples include women discussing abortion in places such as Alabama or Texas where it is now illegal or severely restricted after the 2022 overturning of Roe v.\ Wade; or LGBT+ people who may wish to communicate among each other (e.g., in a work place) but fear the consequences of discovery, especially in countries where it may be illegal, or even in relatively free countries such as the USA where some are now feeling unsafe due to changing societal norms\footnote{\url{https://www.theguardian.com/commentisfree/2023/sep/01/lgbtq-homophobia-rise-new-york}}.

Alice is mainly worried about a curious but honest or semi-honest adversary (stereotyped as `Mallory'). Although Mallory may not be actively looking to find information about Alice, Mallory may discover information that could place Alice in an embarrassing, difficult or dangerous situation. For instance, Mallory could be a family member who disapproves of abortion and discovers Alice's plans via a communication from Alice's supportive friend Bob that shows up on her phone's lock screen, which happens to be lying around on the kitchen table. Mallory could be an IT staff member at a work place who is passively monitoring the WiFi and wired networks, as well as Bluetooth channels for security reasons, thus making it difficult for Alice to set up a handshake with her friends Carol and David to exchange keys for confidential communications about a work place LGBT+ support group.

\noindent\textbf{Alice's capabilities:} We assume that Alice is a `lay' user who may not be technologically sophisticated, and does not have much control over technological equipment (e.g., WiFi routers or CCTVs) in their environment. They may have biometric or similar security over their smartphones but the adversary may be physically close by and could observe the phone through shoulder-surfing and similar attacks, or come across notifications on the lock screen. Moreover, surveillance may be in place to monitor even when the adversary is not present. The ultimate goal for Alice is to establish a secure communication channel with her support group (Carol and David, or Bob) without raising the adversary's suspicion or curiosity. 

 We assume the Alice's primary device for communication is a smartphone, and the available connectivity options are conventional mechanisms such as Cellular, WiFi or Bluetooth. This is attributed to many factors such as purchasing power. The user aims to exchange messages with relatively short duration to convey information privately. 

Alice uses conventional messaging platforms like WhatsApp, iMessage or Signal. However,  Mallory may have physical access to Alice's smartphone when Alice's device is left unattended. We therefore assume that the history of the user's conventional message exchange is known to the adversary. Thus, although these platforms have implemented end-to-end encryption (E2EE) to prevent eavesdropping by third-parties~(including the platform's servers), the adversary's physical access to the survivor's smartphone (e.g., through casually coming across notifications on the lock screen), renders E2EE ineffective. Alice may also innocently install companion messaging apps (e.g., WhatsApp Web) on her work computer without realizing that there is no expectation of privacy from monitoring; thus it is unsafe to simply rely on E2EE, even with self-destructing or scheduled private messages. 

\noindent{\bf{Adversary (Mallory)} and their control over the environment} In our threat model, the adversary Mallory is assumed to have expert knowledge of technology and could have complete control over the conventional communication infrastructure that Alice uses, which may have been set up by the adversary (e.g., the IT staff at he workplace). The adversary is not aiming to actively break Alice's confidentiality, but is curious and will eavesdrop on any private communication of Alice they may come across. This includes message exchange through conventional messaging apps, such as WhatsApp or cellular text messages. They are additionally observant of notifications on the phone, and may thus discover Alice's personal information and activity.

The threat model considers the adversary to be primarily passive with various snooping capabilities in the presence of other people in the environment while actively attempting to restrict Alice when they are alone. However, the adversary is assumed to be technologically capable and can monitor or block conventional channels such as WiFi and Bluetooth. 

In cases when Mallory gets `suspicious' of Alice (e.g., about her sexuality), they may act more maliciously and probe for further information. For example, in some environments, Mallory may be able to monitor audio and visual communications by installing CCTV cameras, enabling the ``drop in'' feature on an Alexa\footnote{https://clario.co/blog/can-you-use-alexa-to-spy-on-someone/}, etc. We assume that Alice is not able to control, dismantle or access any such surveillance systems which may be in place.  Furthermore, the creation of a new communication channel, even with QR code verification on platforms like WhatsApp, can be subverted through tools like QRLJacker\footnote{https://github.com/OWASP/QRLJacking/wiki/QRLJacking-and-Advanced-Real-Life-Attack-Vectors}~\cite[\S3.2.2]{heartfield2018protection}.

\noindent{\bf{Supportive Friend or Network~(Carol)}} Alice has some trusted friends (which we stereotype here as ``Carol'') whom they can contact for help, but a secure channel needs to be setup. Carol can use conventional smartphone messaging applications but is aware that Alice's smartphone contents may be seen by the adversary. Mallory may also have control over Carol's communications environment and have surveillance devices installed for security purposes, for example in work environments shared with Alice.

\section{Keys from sensing shared randomness}
\label{sec:chlngs}
The above threat model sets up the difficulties of creating a secure communications channel between the survivor (Alice) and her supportive friend (Carol) in the presence of adversary Mallory potentially surveilling Alice's network and physical surroundings. A prerequisite for secure communications is to exchange keys. However, as detailed above, Alice may be subject to network as well as audio and visual surveillance. An important contribution of this paper is developing a new method to create a shared secret \textit{without exchanging any data over the network}, by using shared randomness between Alice's and Carol's smartphones. In this section we outline the high level idea and challenges. Sec.~\ref{sec:method} gives the details.

\subsection{Extracting shared randomness across devices without exchanging network data}

The proposed method aims to generate a unique shared secret between smartphones when executed on Alice's and Carol's phones simultaneously. The underlying idea of the method is to combine sensor data (e.g., from gyroscope and/or accelerators) and associated timestamps independently on each smartphone from the series of inertial sensor data and the associated timestamps captured on the smartphones as a result of synchronous random movements of involved smartphones to generate a shared secret.
We explore two mechanisms for generating such synchronous random movements (see Fig.~\ref{fig:basic_idea}): \textit{(i)} placing both smartphones together by holding them with a single hand or \textit{(ii)} placing in the same pocket and moving them together in random synchronous movements. We argue that both of these are natural movements of people with phones, and hence relatively easy to execute over short periods of time without arousing suspicion. 

\subsection{Challenges in synchronising randomness}
\label{sec:sensor_sync_chalngs}
Sensor data on smartphones is a good source of randomness that can be exploited for various purposes such as random number generation \cite{sensor_rand}. However, it is challenging to induce and capture \textit{similar} randomness on \textit{both} devices, for several reasons: 
\begin{enumerate}
    \item The first challenge is the sensitivity of the inertial sensors to the excitation trigger. Carlsson et al.~\cite{9458231} and He et al.~\cite{10227897} show that the measurement error variance between two inertial sensors can be around $10\%$. 
    \item The positional sensors, such as the gyroscope, measure the quantities \textit{``with respect to''} the positional axis. The misalignment of these axes between two devices significantly changes the gyroscope readings. 
    \item The gyroscope sensor (similarly other sensors) is also prone to drift in sensor data over time \cite{sensor_drift}. 
    \item Finally, the sensor data is also subject to manufacturer-dependent provisions. For example, Jan et al.~\cite{7512040} show that the gyroscope data across different smartphone models, possibly using gyroscope hardware from different manufacturers, tends to agree with each other over certain confidence levels only. 
\end{enumerate}

The next section shows how to systematically overcome these challenges to generate secrets on two smartphones using gyroscope data independently. 

\section{Secret Generation Method}
\label{sec:method}
\begin{figure*}[t]
  \centering
  \includegraphics[width=\textwidth]{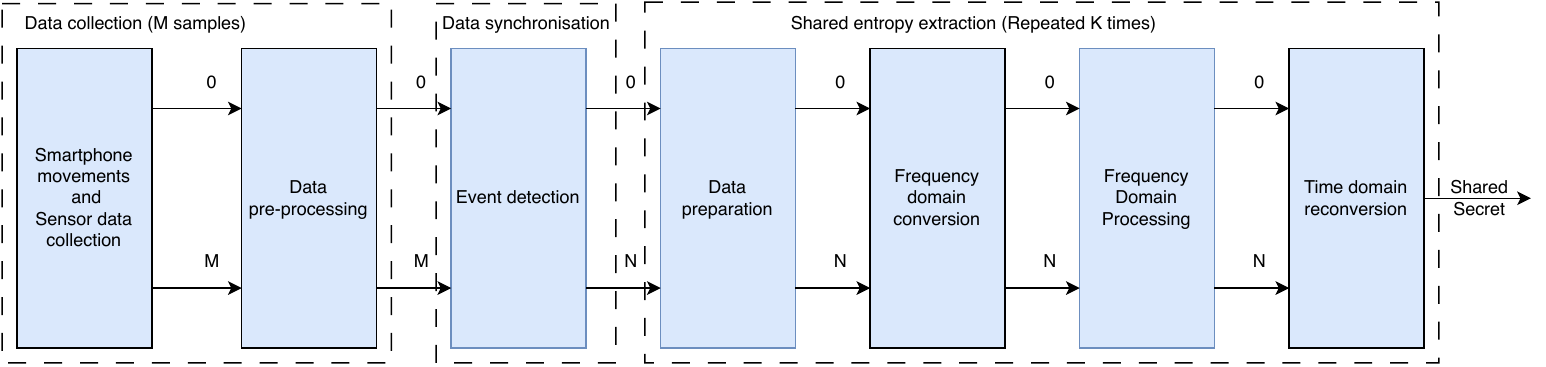}
  \caption{Shared secret generation method}
  \label{fig:method}
\end{figure*}

The smartphone inertial sensors capture all the random movements that a device may be subjected to. However, we need to align the sensed movements across two devices. We develop on-device processing to first detect an initial data alignment point between two smartphones independently on each smartphone. Each device then converts the sensor data and associated timestamps after the alignment point into a shared secret by extracting the shared randomness by first representing the sensor data and associated timestamps as a discrete-time complex signal and then applying a DFT-based method to reduce the variations in the signal power. The method combines the usage of multiple sensor data variants and frequency spectrum adjustment to generate a longer shared secret with higher entropy.

Fig~\ref{fig:method} depicts the end-to-end process on a single smartphone involved in generating a shared secret. The figure shows that the process is divided into three phases: data collection (\S\ref{sec:sensor_data_gen}), data synchronisation (\S\ref{sec:data_sync}), and shared randomness extraction (\S\ref{sec:sh_ent_extract}).

\subsection{Data collection}
\label{sec:sensor_data_gen}
Modern smartphones typically have two inertial sensors: an accelerometer (capturing linear movements) and a gyroscope (capturing angular movements). In the current version of the method, we propose using the smartphone gyroscope to capture these movements in digital format. The gyroscope shows better robustness under dynamic conditions such as vibrations \cite{10147912}. Using other sensors as an entropy source is one of the future directions for the method. 

As the method aims to generate a shared secret by extracting similar randomness between more than one device, the gyroscope sensor will require a suitable mechanism to align the source reference across smartphones, i.e. principal axes of inertia (cf. Fig.~\ref{fig:basic_idea}). The method requires users to align principal axes before starting the secret generation process. This can be done by aligning the phones in the user's hand, or placing them in the same pocket. No user interruption is necessary after reference point alignment. 

The gyroscope then captures the synchronous random movement of both smartphones. Both smartphones collect gyroscope data for the same predefined time interval (approx 1 min or 150 samples), whose duration is selected based on empirically observed probability of successful secret generation, stable shared entropy and user convenience. If ${x_t, y_t, z_t}$ is a gyroscope data triplet recorded at time $t$, then the gyroscope data at same time $t$ used for further entropy extraction is given by
\begin{equation}
    x_g(t) = x_t * sin(\theta/2)
\end{equation}
The value of $\theta$ is computed as per the Android motion sensor API description\footnote{\url{https://developer.android.com/develop/sensors-and-location/sensors/sensors\_motion}}. This design decision is based on the experimental evaluation that results in a higher success rate. 

Due to the inherent time offset between sensor data collection on both smartphones, the sensor data must be synchronised to generate the shared entropy. The data synchronisation algorithm performs this task as described in Sec.~\ref{sec:data_sync}. However, our experiments show that raw timestamps (at millisecond granularity) and sensor data fail to produce the desired synchronisation point due to the sensor data-specific event detection method employed instead of external physical event detection. The proposed sensor data-specific event detection method detects the synchronisation point against a specific property of the entire data, such as the average value. Hence, the timestamps are appropriately scaled before alignment to improve the shared secret/entropy accuracy. Similarly, gyroscope data is scaled suitably to incorporate its contribution in the final shared entropy. Specifically, the gyroscope data is scaled as per Eq.~\ref{eq:gdata_scal} while embedding entire data-specific shared randomness. If $X_g = x_g(t) |_{\forall t}$ is gyroscope data then scaled data is given by $X'_g(t) = x'_g(t)|_{\forall t}$ where 
\begin{equation}
\label{eq:gdata_scal}
    x'_g(t) = x_g(t) * Var(X_g) * K_g
\end{equation}
The value of $K_g$ is chosen empirically, leading to accurate synchronisation and a higher success rate. Further, the sensor data is smoothened to reduce noise variations in the sensor data. Specifically, the commonly used exponentially weighted moving average (EWMA) method is used that combines the sensor data value at a discrete time ``\textit{k}" with previous data using $x[k] = \alpha.x[k] + (1-\alpha.x[k-1])$. The linear operation in EWMA does not affect the inherent randomness properties of sensor data while averaging the effect of smartphone movement spread across two consecutive data points.
The data collection phase collects and prepares the series of captured gyroscope data and associated timestamps for further processing to generate a shared secret. 

\subsection{Data synchronisation}
\label{sec:data_sync}
Due to human limitations in synchronising the gyroscope data collection at the necessary time granularity, there is a time offset in the gyroscope data between both smartphones. The ``\textit{Data synchronisation}" and ``\textit{Shared entropy extraction}" (refer Sec~\ref{sec:sh_ent_extract}) are two independent phases of the proposed shared secret generation method shown in Fig.~\ref{fig:method}. However, ``\textit{Shared entropy extraction}" phase enables two smartphones to start measuring gyroscope data at the same time. The considered threat model (refer Sec.~\ref{sec:th_model}) describes the attacker being equipped with a surveillance system that can observe any specific movements.

To estimate the synchronisation point in the presence of a video surveillance system, the method applies the ``sensor-data-specific-bound-based" threshold detection technique to independently establish that point as a synchronisation point on both smartphones for further shared secret extraction. This removes the necessity for externally induced specific movement that can be served as synchronisation point. The dependency on the sensor data-specific bound instead of the external physical event adds randomness in selecting sensor data points for secret generation thus making it resilience against the video surveillance. 

\begin{figure}[t]
  \centering
  \includegraphics[width=\linewidth]{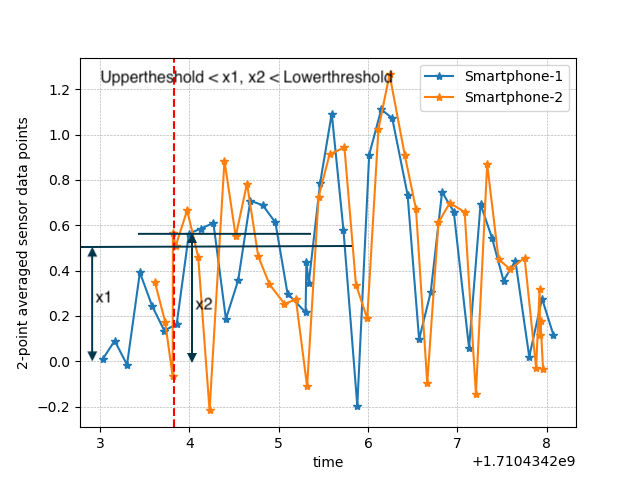}
  \caption{Synchronised data after common feature detected}
  \label{fig:sync_data}
\end{figure}

Fig.~\ref{fig:sync_data} shows the threshold detection idea behind the employed event detection algorithm over the synchronised smartphone gyroscope data. Let the angular velocity step at discrete time instance $n$ be denoted $\Delta x_n$. The method defines two thresholds, namely $Xth_{lower}$ and $Xth_{upper}$. Different orchestration of smartphone movements produces angular velocities of different magnitudes. That implies the detection threshold cannot be fixed or constant. To overcome this problem, the method calculates these thresholds as a function of collected sensor data. It is a variant of the entropy-based thresholding method \cite{7839934} with two steps and step size associated with specific percentile value from the given sensor data series. Specifically, Eq.~\ref{eq:th_vals} gives the formulation of these thresholds. 
\begin{eqnarray}\label{eq:th_vals}
    Xth_{upper} &=& 75\%le\ sensor\ data\ value    \\  
    Xth_{upper} &=& Xth_{upper}(1+K/100)    
\end{eqnarray}
$K$ is a configurable variable we have empirically selected through calibration with smartphone logs. The feature is said to be detected at discrete time $n$ if the $\Delta x_n$ satisfies Eq.~\ref{eq:th_detect}.
\begin{equation}\label{eq:th_detect}
    Xth_{lower} < \Delta x_n < Xth_{upper}
\end{equation}

Both smartphones perform the data synchronisation independently. However, synchronous movements of both smartphones during the data collection phase and the application of EWMA ensure that they synchronise on similar data points. Fig.~\ref{fig:sync_data} also shows the snippet of the synchronised data after applying the feature detection technique. Though the movement of both smartphones is synchronised and axes are aligned before the movement orchestration, the inherent offset between sensor values and non-identical time-slot reference for sensor data recording produces ``approximately" similar delta angular velocities on both smartphones as discussed in Sec.~\ref{sec:chlngs}. 

The method discards the data before the synchronisation point to extract the shared entropy.

\subsection{Extracting shared randomness}
\label{sec:sh_ent_extract}

As discussed in the previous section, the data synchronisation phase identifies the discrete time instance at which both smartphones experience approximately similar delta angular velocity. However, the inherent discrepancy of data points prohibits the direct usage of the detected synchronisation point as the shared entropy (refer Sec.~\ref{sec:sensor_sync_chalngs}). Hence, the method extracts the shared randomness from the consecutive $N$ data points after the detected synchronisation point. $N$ is a function of success rate, entropy addition and user convenience. Reminder of section explains steps involved in shared randomness extraction process.

\paragraph{\bf{Data preparation}}
By definition, the gyroscope angular velocity and timestamp are independent signals. The gyroscope angular velocity measurement data and associated timestamps form two independent data series or signals. Hence, the method combines these two independent signals as a single complex signal by considering gyroscope data as the \textit{real} part and timestamp as the \textit{imaginary} part of the complex signal. Combining data in this way parallelises the processing necessary to extract the entropy from each series and combine it. Specifically, the method combines the sensor data with the associated timestamp using Eq.~\ref{eq:data_combine}.
\begin{eqnarray}
\label{eq:data_combine}
    x[i] = s[i] + j ts[i]
\end{eqnarray}

\paragraph{\bf{Frequency domain conversion}} 
The sensor data, the primary source of entropy, results from synchronous random movements of both smartphones. The hypothesis is that the synchronous random movements of smartphones generate angular velocities on both smartphones with frequencies shared across smartphones. 

The Discrete Fourier Transform (DFT) is a commonly used method to retrieve the frequency spectrum of the given discrete time signal. The DFT combines the gyroscope data and associated timestamps through efficient multiplication. The Inverse DFT ensures the error-free transition back to the original data domain (i.e. angular velocity and time). The method uses the Fast Fourier Transform (FFT) algorithm to compute the DFT and retrieve frequency components of captured angular velocities. The FFT algorithm efficiently implements Discrete Time Fourier Transform for data size of length $2^n$ where $n=0,1,2,...,n$. The Fourier transform extracts the contribution of each input data value on a given set of fundamental frequencies or harmonics limited by the data length $N$. This data transformation is represented by Eq.~\ref{eq:dft}
\begin{eqnarray}
\label{eq:dft}
    X_k = \sum_{m=0}^{N-1}x_me^{-j2\pi km/N}
\end{eqnarray}
It can be seen from Eq.~\ref{eq:dft} that the parameter $N$ controls the total number of frequencies in the resulting frequency spectrum. 

\begin{figure}[t]
  \centering
  \includegraphics[width=\linewidth]{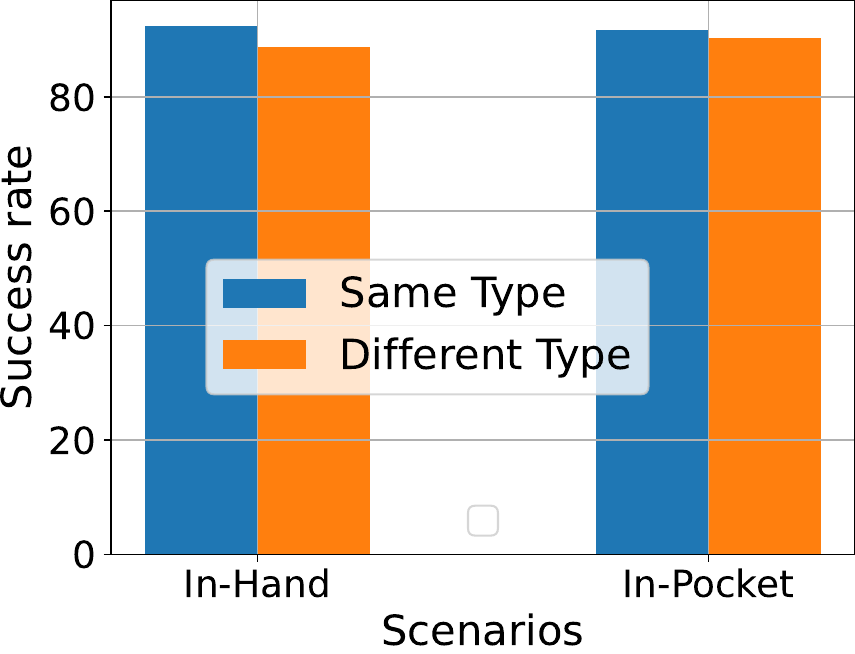}
  \caption{Success rate}
  \label{fig:succ_rate}
\end{figure}
\begin{figure}[ht]
    \centering
    \includegraphics[width=\linewidth]{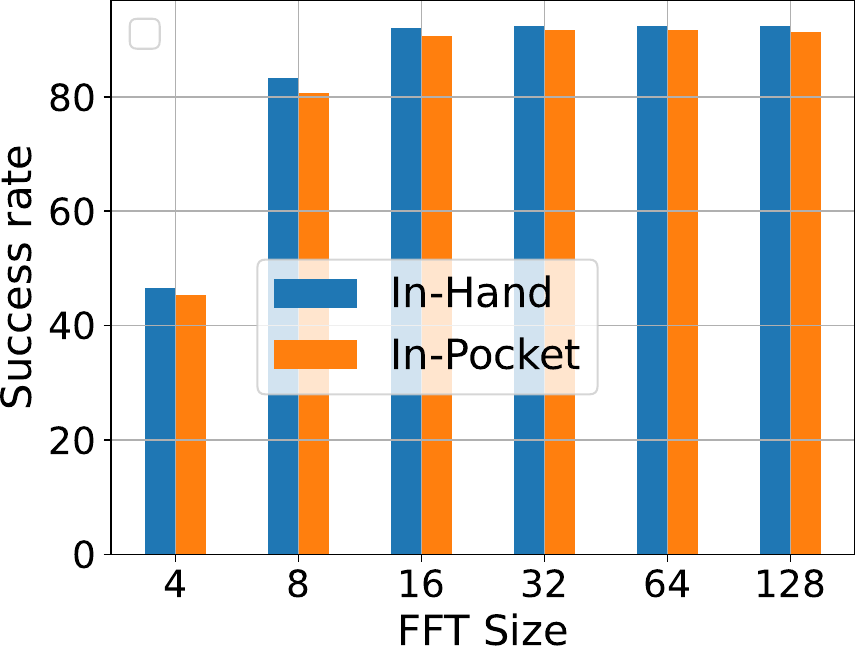}
    \caption{Effect of FFT Size on the success rate for the same type of smartphones}
    \label{fig:succ_rate_fftsizes_sm}
\end{figure}

\paragraph{\bf{Frequency domain processing}}
\label{sec:freq_process}
The method uses the \textit{``Maximum Entropy"} \cite{1524863fftspectralanalysis} approach to get the maximally random (maximum entropy) data point within the set of input data points (sensor data and timestamp) range. It requires modifying the underlying frequency spectrum to adhere to conditions specified for \textit{``Maximum Entropy"} approach.

The $X_k[0]$ or the $0^{th}$ frequency component is called DC bias \cite{4762004}. It represents the average input data values or the smartphone's movements represented by sensor data. Higher frequencies represent the more dynamic parts of the input as frequency increases. Every orchestration of smartphones is associated with DC bias specific to orchestrated movement. Thus, it forms the fingerprint of specific smartphone movements at specific instances of time. One approach would be to extract this shared DC bias as a new frequency spectrum for shared entropy. However, this approach filters out more random data from the computation, which is undesirable in entropy extraction processing. 

To overcome this issue and to add the necessary randomness while generating identical secrets on both smartphones, the proposed method computes the frequency spectrum as in Eq.~\ref{eq:mod_freq_spec}
\begin{eqnarray}\label{eq:mod_freq_spec}
    E = \sum_{k=0}^{N-1} X_k \\
    X^{'}[k] = g(k)*E
\end{eqnarray}

The $X^{'}_k$ preserves the randomness present in the original data on specific frequencies, a multiplication of combined randomness and a multiplication factor represented by $g(k)$. This assignment distributes the combined randomness on all frequencies. However, it raises the question of creating substantially different copies of combined randomness to generate substantially different contributions to the final secret by given frequency. Very close values for frequency components will produce output at a single or few time instances due to the impulse response property of DFT. This leads to having a long train of ``0"s appearing in the secret, which is undesirable from a randomness perspective. The method proposes to use the $g(n) = n$ function through empirical analysis. 

\paragraph{\bf{Time-domain reconversion}}
\label{sec:time_domain_reconversion}
The Inverse Discrete Fourier Transform (IDFT) is a complementary method to convert the frequency spectrum into time domain data. The IDFT is computed using Eq.~\ref{eq:idft}. 
\begin{eqnarray}
\label{eq:idft}
    X_m = \frac{1}{N}\sum_{k=0}^{N-1}x_ke^{j2\pi km/N}
\end{eqnarray}
The proposed method applies the Inverse Fast Fourier Transform (IFFT) algorithm on the frequency spectrum constructed in Eq.~\ref{eq:mod_freq_spec} for back conversion to time domain data and extracts similar shared entropy. The proposed method computes the magnitude of computed time domain data points using Eq.~\ref{eq:power}. Further, it is rounded to the nearest integer value. This process filters out the effect of the difference between sensor data on both smartphones as it is adjusted to an appropriate value satisfying the Eq.~\ref{eq:att_noise_margin}. 
\begin{eqnarray}\label{eq:att_noise_margin}
\label{eq:power}
    x &=& a + j b\\
    Magnitude(x) &=& \sqrt{{a}^2 + {b}^2} 
\end{eqnarray}
As the input data is significantly biased with timestamp due to necessary scaling, the few least significant bits capture the gyroscope data-based randomness in the computed data magnitude. A single bit of each reconverted data point's magnitude is extracted as a shared entropy. If $b_k$ represents the \textit{LSB} of the data magnitude at point $k$, then sequence $${b_0b_1b_2......b_n }$$ represents the extracted shared entropy.

To extract the additional entropy from the input gyroscope data, the method repeatedly applies frequency domain conversion, frequency domain processing, and time-domain conversion to multiple variants of input data. These variants follow the Eq.~\ref{eq:data_var} generated using the original input data in Eq.~\ref{eq:data_combine}.
\begin{eqnarray}
\label{eq:data_var}
    x[i] = f(k) * s[i] + j ts[i]
\end{eqnarray}
where $k$ represents the interaction index. The function $f(x)$ solves the problem of creating multiple copies of input data. It should overcome the challenge of lowering the gyroscope data deviation across smartphones while creating enhanced shared entropy by each newly created input data variant. The method currently proposes to use $f(n)=1+\frac{n}{10}$ function, which is a low slope function varying the data in smaller steps but substantially different in terms of digits in the input data. Each interaction generates the sequence of extracted bits. The method concatenates bit sequences computed by all interactions in a First-Come-Fisrt-Serve order to generate a final extracted shared secret/entropy. 

\renewcommand{\algorithmicrequire}{\textbf{Input:}}
\renewcommand{\algorithmicensure}{\textbf{Output:}}

\begin{algorithm}
\caption{Shared entropy extraction}\label{algo:sh_ent_ext}
\algorithmicrequire {Valid gyroscope data $S =[s_0, s_1,...,s_n]$ and associated timestamps $T=[ts_0, ts_1,...,ts_n]$} \\
\algorithmicensure {Shared Entropy \textbf{SE}}
\begin{algorithmic}[1]
    \State $T[i] = T[i] * T_{scale}$ ; $T_{scale}$ empirically calibrated
    \State $S[i] = S[i] * Var(S)*C$ $where\ C=10000; calibration$
    \State \textbf{Synchronise:} $S \rightarrow S_a=[s_{a0}, s_{a1},...,s_{an}]$ ; $T \rightarrow T_a=[ts_{a0}, ts_{a1},...,ts_{an}]$ 
    \State \textbf{Initialise:} $ N=FFT\_size, f_k =\ f(k) = 1+\frac{k}{10}$ , $k = num\_factors (``calibrated\ value")$
    \For {$m = 1$, $m \leq num\_factors$, $m{+}{+}$}
        \State $x_i = f_m * S_a[i] + j T_a[i]; X=[x_0, x_1,...,x_N]$
        \State Transform $X$ to frequency domain using DFT, $X_k = \mathcal{F}(X)$
        \State $E = \sum_{k=0}^{N-1} X_k$
        \State $X^{'}[k] = g(k)*E$
        \State Retransform $X^{'}_{k}$ to time-domain using IDFT; $X^{'} = \mathcal{F^{-}}(X^{'}_{k})$; $X^{'}(n) = a + jb$
        \For {$j = 1$, $j \leq FFT\_Size$, $j{+}{+}$}
            \State $E^{'}_j = Round(\sqrt{{a}^2 + {b}^2})$
            \State $SE = SE || LSB(E^{'}_j)$
        \EndFor
    \EndFor
    \State $return\ SE$ 
\end{algorithmic}
\end{algorithm}
The Algo.~\ref{algo:sh_ent_ext} describes the flow of the proposed secret generation method. 

\section{Evaluation}
\label{sec:eval}
We evaluated the proposed secret generation method for its accuracy, security attributes, and robustness using the Android app based experimental setup.

\subsection{Experimental setup}
The experimental setup consists of an Android app developed to capture the gyroscope sensor data and associated timestamp on the device. This app is loaded onto two different smartphones. The experiment uses two Google Pixel 7s and one Samsung Galaxy A53 5G smartphone. The experiments consists of two types of scenarios for the smartphone movements, viz In-hand, In-Pocket.  

\paragraph{\bf{Experiment Data collection}}
We implemented a test app on the Android operating system to understand the practical considerations and evaluate the various performance metrics. The developed test app is supported on the Android API version 32 or Android OS version 12, i.e. Android Snow Cone. The app does not need any special permission from the user. It uses the JTransforms \footnote{``com.github.wendykierp:JTransforms:3.1" dependency} Java library to implement the Fast Fourier Transform. The app is currently distributable as a private app solely for method calibration/evaluation purposes. We used off-the-shelf Google Pixel-7 smartphones to evaluate the performance of the proposed method on the same type of smartphones. We also used a Samsung Galaxy A53 smartphone to add diversity to our results. Each smartphone collects $150$ data points when triggered before performing certain synchronous movement.  This leads to generation of two data logs, one on each smartphone, as a result of given movement. The set of two of these logs forms one experiment. The evaluation uses $300$ experiments for each type of movement. Unless mentioned otherwise, it uses $FFT\ Size=64$ and number of input variants (i.e. $factor$ variable in Algo.~\ref{algo:sh_ent_ext}) as $5$. 

\begin{figure}[ht]
    \centering
    \includegraphics[width=\linewidth]{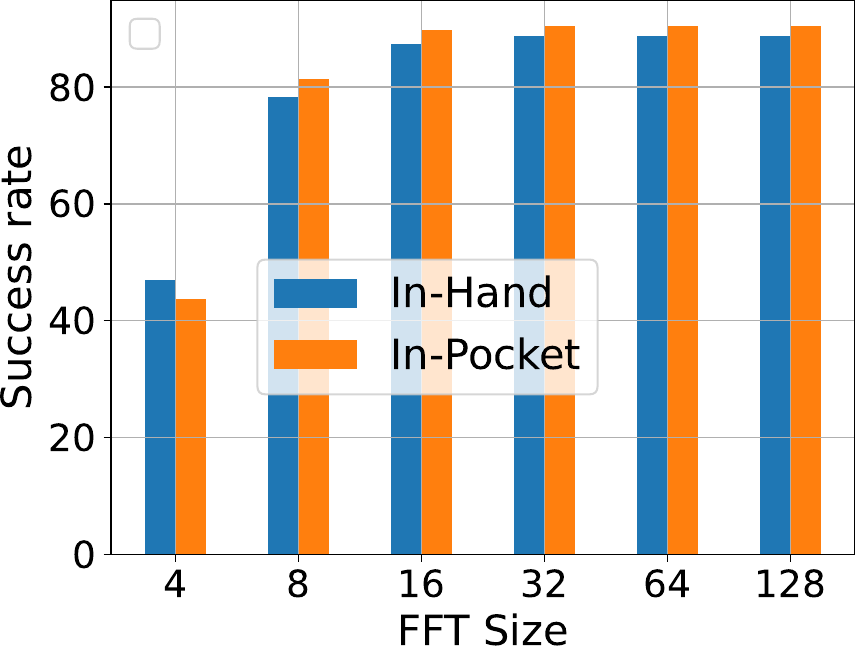}
    \caption{Effect of FFT Size on the success rate for different types of smartphones}
    \label{fig:succ_rate_fftsizes_dm}
\end{figure}

Fig.~\ref{fig:succ_rate} shows the success rate, i.e., the probability that the logs on the two smartphones have arrived the same shared secret. 
We achieve a success rate of over 90\% when the phones are in the user's hand, regardless of whether the phones are of the same type or of different types (i.e., the inertial sensors and firmware are identical or potentially different). However, when the phones are in the pocket, the success rates drop slightly, achieving 86--91\% accuracy depending on whether different or similar phone types are used.  This experiment uses the $FFT\ Size=64$. We next examine sensitivity to the size of the FFT.

\subsection{Effect of FFT Size on success rate} Fig.~\ref{fig:succ_rate_fftsizes_sm} and \ref{fig:succ_rate_fftsizes_dm} shows the effect of \textit{FFT Size} on the success rate when using the same type or different type of phones respectively. With smaller FFT size, the success rate decreases due to inherent variations in the gyroscope data across smartphones. However, with an FFT size of 16 or over, we achieve similar success rates; hence, we use a FFT size of 16.

Our experimental results also shows that the variations in $num\_factors$ do not affect the success rate for given selection of function $f(x)$ (see Algo.~\ref{algo:sh_ent_ext}). 

\begin{figure}[ht]
    \centering
    \includegraphics[width=\linewidth]{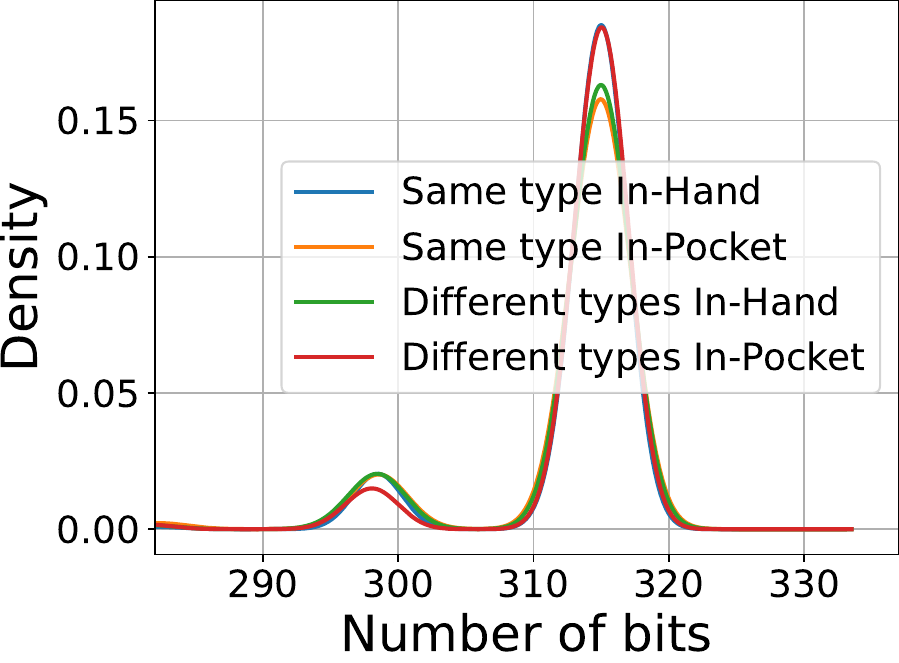}
    \caption{Distribution of lengths of secrets}
    \label{fig:sec_len}
\end{figure}

\subsection{Effect of timestamp scaling on success rate}
The timestamp scaling divides the entire gyroscope data into multiple slots depending on the chosen scaling factor. The gyroscope data timestamps are recorded at millisecond granularity. Hence, the scaling factor of (1/1000) divides the approx. $1$ minute of data into approx. $60$ slots, with each sample within each having the same timestamp. We have evaluated the effect of varying the timestamp scaling factor, resulting in different numbers of time slots. 
\begin{figure}[ht]
    \centering
    \includegraphics[width=\linewidth]{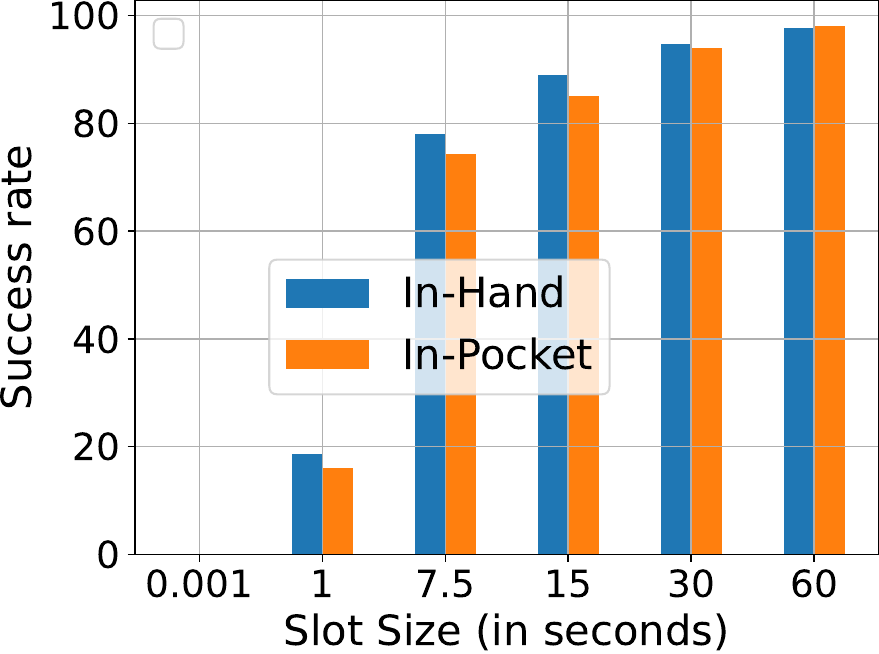}
    \caption{Effect of timestamp scaling on the success rate for the same types of smartphones}
    \label{fig:succ_rate_tscal_sm}
\end{figure}

Increasing the number of slots through the scaling factor decreases the number of samples per slot. Thus reducing the probability of a specific data sample (satisfying the threshold detection criterion) falling into the slot with the same scaled timestamp on both smartphones. This is also sensitive to the offset between both smartphones' gyroscope data recording starting times, resulting in different synchronisation thresholds between smartphones. Larger time slots normalise this difference, thus reducing the effective offset between scaled timestamp values. Our results, shown in Fig.~\ref{fig:succ_rate_tscal_sm}, depict a similar observation. As the slot size increases, more samples fall into slots with similar scaled timestamp values and approximately similar normalised detection thresholds. Hence increasing the success rate.    

\subsection{Size of the secret}
Fig.~\ref{fig:sec_len} shows the bit size distributions of the generated secrets for $FFT\ Size = 64$. The expected maximum secret length for a given \textit{FFT size} is equal to $FFT\ size\ *\ num\_factors$ (e.g. 64 x 5 = 320) as per Algo.~\ref{algo:sh_ent_ext}. In practice we obtain variable length secrets with the bitsize density peaking near $300$ and $315$. This result is robust regardless of whether the phones are of the same type and whether we hold the phones in the hand or place them in pockets. This suggests that the generated secrets are long enough to withstand several common types of guessing attacks.

\subsection{Scalability analysis}
\begin{figure}[ht]
    \centering
    \includegraphics[width=\linewidth]{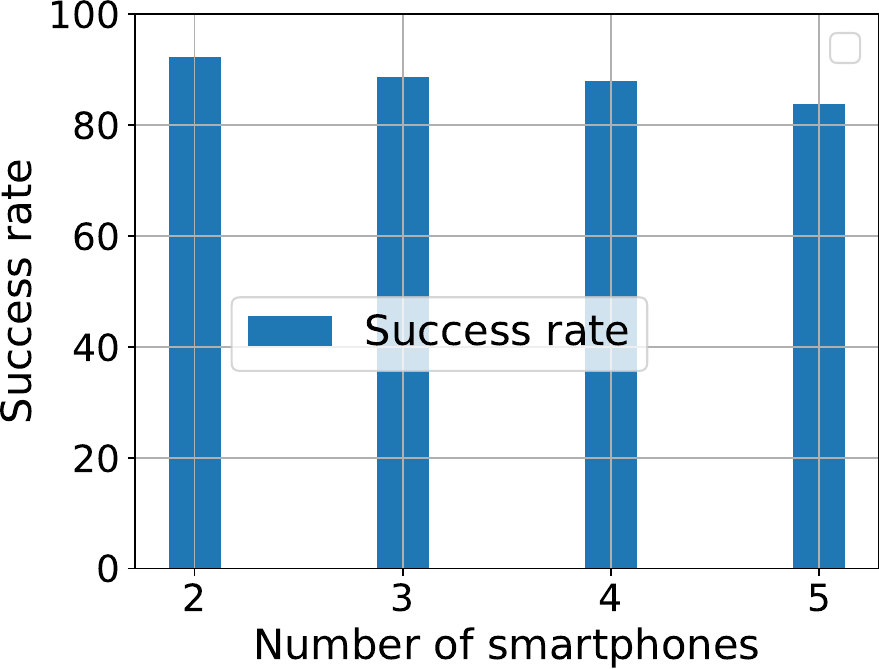}
    \caption{Variation in success rate with number of smartphone involved}
    \label{fig:succ_rate_scal}
\end{figure}
The method runs the synchronisation and shared secret extraction procedures independently on each smartphone. However, to understand the effect of involving multiple smartphones on the shared secret generation, we performed experiments with 3, 4, and 5 smartphones. We performed 150 experiments with each combination of smartphones (Google Pixel 7, Google Pixel Pro, Samsung Galaxy, and One Plus models). The experiment provides 15 seconds of initial setup time, to allow pressing the start button and putting all smartphones together in the pocket. We considered the scenario of smartphone movement while keeping smartphones in the pocket.

As shown in Fig.~\ref{fig:succ_rate_scal}, the success rate remains above $80\%$ for all smartphone combinations. However, there is a decrease in the success rate from $90\%$ for two smartphone cases to $84\%$ for five smartphone cases. As the number of smartphones increases, the similar alignment of all smartphones is challenging to maintain. This leads to the wrong calculation of synchronisation points based on recorded offseted gyroscope readings. Such gyroscope reading variations are amplified by the method's inherent mechanism of using multiple scaled values of gyroscope data. Moreover, the time-offsetted start of gyroscope measurement on different smartphone can led to one smartphone detecting the synchronisation point in the timestamp zone when one of the another smartphones hasn't started recording the data. Hence, we recommend to induce smaller movements in the beginning for few seconds (equal to time duration between starting measurement on first and the last smartphone, typically 5 seconds). The combined effect of wrong synchronisation point calculation and gyroscope data offset implication lowers the success rate when more smartphones are involved. 

\subsection{Understanding errors}
\begin{figure}[ht]
    \centering
    \includegraphics[width=\linewidth]{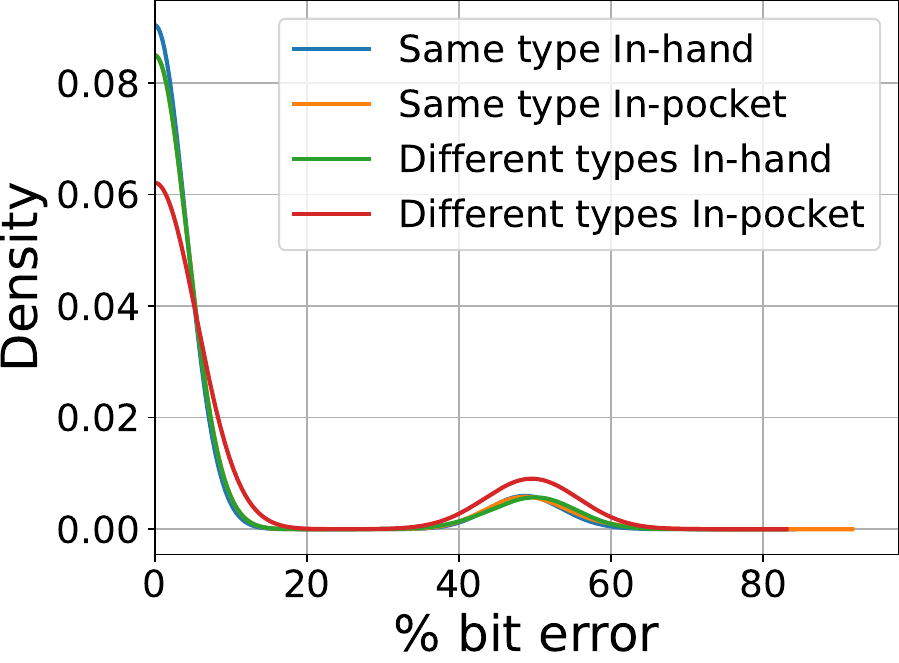}
    \caption{Distribution of number of bits in error for non-identical secret generation on both smartphones}
    \label{fig:error_dist}
\end{figure}

Finally, we examine the cases when the smartphones are \textit{not} able to agree on a shared secret, and examine how many bits differ.  
Fig.~\ref{fig:error_dist} shows the distribution of the percentage number of bits that differ between two secrets when the non-identical secrets are generated on both smartphones in an experiment.
In more than $90\%$ of cases, fewer than $20\%$ of bits differ. Such errors may be fixed by letting the synchronisation run for a slightly longer duration, or by adding an error correcting code. In our implementation (\S\ref{sec:securewhispersapp}), we let the synchronisation run and inform the user with a brief vibration when the shared secret has been generated.

\subsection{Entropy and Randomness}
\begin{table}[htbp]
    \centering
    \caption{NIST 800-22 test results}
    \label{tab:nist-800-22}
    \begin{tabular}{ |c|c|c|}
    \hline
    \multicolumn{3}{|c|}{NIST 800-22 test results} \\
    \hline
    Test & p-value & \% Passing \\
    \hline
    Frequency  & 0.50 & 99.34 \\
    Cumulative sum & 0.52 &  99.34 \\
    Runs & 0.50 & 99.34  \\
    Linear complexity & 0.73 &  93.59 \\
    Block frequency & 0.50 & 98.98  \\
    Longest run & 0.50 &  99.16 \\
    FFT & 0.51 &  99.65 \\
    Serial & 0.52 &  99.30 \\
    \hline
    \end{tabular}
\end{table}

\begin{table}[htbp]
\centering
\caption{Computed Entropy}
\label{tab:comp_entropy}
\begin{tabular}{ |c|c|c| } 
\hline
Smartphones & Scenario & Entropy (/8 bits) \\
\hline
\multirow{2}{6em}{Same Type} & In-hand & 6.12 \\ 
& In-pocket & 6.20 \\ 
\hline
\multirow{2}{6em}{Different Types} & In-hand &  6.40 \\ 
& In-pocket & 6.15  \\ 
\hline
\end{tabular}
\end{table}

Next we examine how secure the generated passwords are, by executing the NIST 800-22 tests \footnote{https://csrc.nist.gov/pubs/sp/800/22/r1/upd1/final} on the final shared secret generated by the proposed method. This test suite performs various tests, such as frequency and block tests, to check the randomness of the given bit sequences. It reports the decision on whether the given bit sequence passes the randomness test through $p-value$. The threshold for passing these tests is $0.001$. If $p-value>0.001$, the given bit sequence passes the test. Table~\ref{tab:nist-800-22} shows the results of NIST 800-22 tests for the proposed method's final secrets. Since all test results report more than $0.001$ for more than $99\%$ of secrets, the generated shared secrets pass the NIST 800-22 randomness tests. 

We further investigated the entropy of the generated secrets to verify their usability as a seed for encryption key generation. We followed the NIST 800-90B \footnote{https://github.com/usnistgov/SP800-90B\_EntropyAssessment} standard to compute the entropy. For each input bit sequence, the test suite computes the entropy of each 8-bit block. Table~\ref{tab:comp_entropy} shows the entropy of secrets for each type of scenario computed using the NIST 800-90B test suite. The entropy in bits of all secrets is 6.632 bits per byte against 7.7 bits per byte \cite{10.1145/3442520.3442528} entropy of the raw gyroscope data. Approximately 1-bit entropy is lost as non-shared entropy between smartphones. Table~\ref{tab:entropy_compare} shows that this entropy is comparable to the two methods of entropy generation used in OpenSSL~\cite{8939840}.

\begin{table}
\centering
\caption{Entropy comparison}
\label{tab:entropy_compare}
\begin{tabular}{ |c|c| } 
\hline
Method & Entropy (/8 bits)  \\
\hline
/dev/urandom (Linux-Ubuntu 16.04 LTS) \cite{8939840}& 7.86--7.88 \\
\hline
OpenSSL-RTC\footnote{Real Time Clock} Linux-Ubuntu 16.04 LTS   \cite{8939840} & 6.59--6.65 \\
\hline
Proposed method & 6.1--6.6 \\
\hline
\end{tabular}
\end{table}

\begin{figure}[ht]
    \centering
    \includegraphics[width=\linewidth]{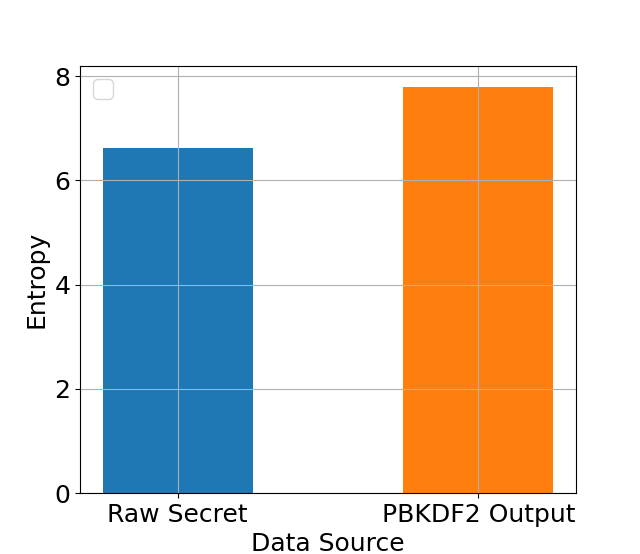}
    \caption{Improvement in entropy in bits/8 bits}
    \label{fig:entropy-improvement}
\end{figure}

\section{SecureWhispers App}
\label{sec:securewhispersapp}
\begin{figure}[t]
  \centering
  \includegraphics[scale=0.45]{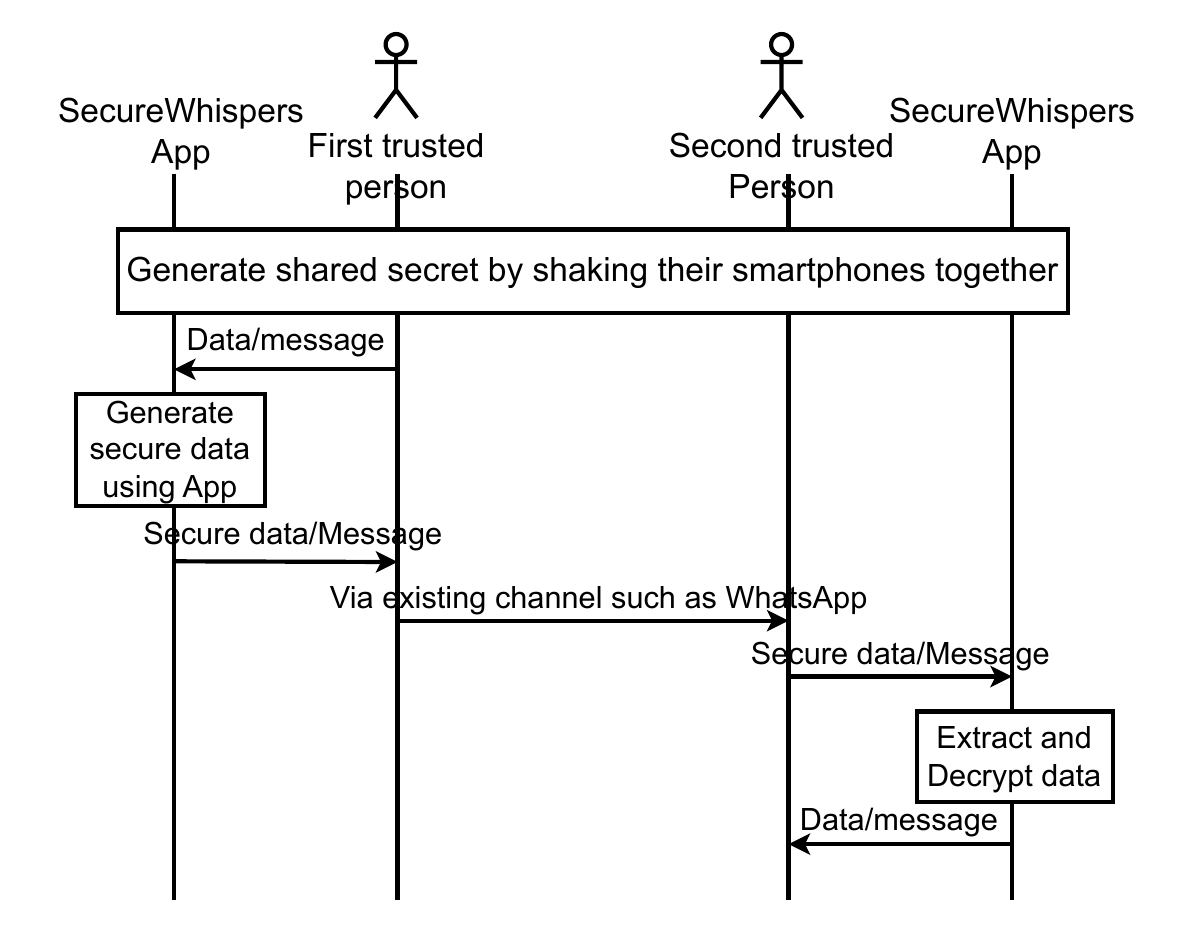}
  \caption{App's illustrative usage scenario}
  \label{fig:app_usage}
\end{figure}

We develop ``SecureWhispers" (refer Appendix~\ref{sec:appdx_swa}), an Android application enabling users to create a secret channel within normal conversations. Fig.~\ref{fig:app_usage} shows the typical usage scenario of ``SecureWhispers" App. 

\subsection{Shared Secret Generation}
To generate a shared secret between users, both users simultaneously select the \texttt{START} button beneath ``Generate Secret'', and are then provided a few seconds to put the phones together to begin the secret generation. The phones vibrate briefly to provide user feedback when they have finished converging on a shared secret.

\subsection{Encrypting Messages}
Once the shared secret has been generated, it is hashed using PBKDF2 with HMAC-SHA256 to create a shared 256-bit encryption key, which is then used to encrypt messages under AES-256-CBC with a secure random IV. Fig.~\ref{fig:entropy-improvement} shows the impact of this on entropy.

We choose these algorithms as they are well-established, standardised cryptographic methods that are secure against a much more technical adversary than is expected from our established threat model in Section~\ref{sec:th_model}.

\subsection{Security Analysis}
We now briefly discuss possible attack scenarios that can lead to a secret compromise in SecureWhispers, based on the threat assumptions in Section~\ref{sec:th_model}. 
The adversary can deploy various attacks. At its simplest, the adversary can guess the secret. For security analysis, our baseline assumption is that Mallory is aware such a solution exists for secure messaging with the supportive friend. In this section, we discuss the attack vectors to our designed scheme.

\paragraph{\bf{Attacks to the Key}}
Our experiments show that the generated key has a variable length in the range of 290 to 320 bits following a bell-shaped distribution~(see Figure~\ref{fig:sec_len}). For Mallory, the first challenge is to guess the length of the established key. Moreover, since the key generation is executed with no communication across smartphones, the adversary cannot execute eavesdropping or man--in--the--middle attacks to acquire the key over the wire.

The generated secret is 256 bits long. Evaluating the outputs with the NIST 800-22 test suite, we achieve 7.798 bits/byte of entropy, providing $256 \times 7.798 \div 8 = 249.5$ bits of entropy in the full key. This implies that our proposed scheme can withstand brute-force attacks on the key, which have a minimum search space of $2^{249}$.

The adversary may look into sources of entropy to bypass guessing the keys. This would require Mallory to have access to the exact movements of Gyroscope and timestamps. This is not an unrealistic assumption as Mallory may have a surveillance system in place to observe the movements of Alice and Carol. However, this is a challenging task as it requires an intrusive attempt to recognise the exact start and end of the key generation process. In our scenario, this happens when both smartphones are hidden in Carol's pocket while walking around. Thus, to acquire the exact Gyroscope data, Mallory should mimic the exact movements of Carol. This, in addition to the added inherent noise in the Gyroscope sensor reading, would make the replication of Gyroscope data infeasible.


The threat model considers the adversary equipped with video surveillance capabilities. This provision opens the possibility of ``\textit{Mimicry attack}" executed as an adversary trying to steal the shared secret by ``\textit{mimicking}" the smartphone movements. The secret generation starting point knowledge is crucial for the successful shared secret generation. Our results show that the adversary can successfully regenerate the shared secret $80\%$ of time if it knows the start point at the granularity of the time slot defined by the timestamp scaling factor. Thus, under the starting point secrecy conditions, such as keeping smartphones in pockets or bags/purses, the proposed method generated cryptographically strong and extremely private shared secrets. 

\paragraph{\bf{Attacks to the Secure Channel}}
The adversary can impersonate the survivor to supportive friend (Carol). Alice and Carol communicate through conventional messaging services, using the encrypted messages. Thus, SecureWhispers partially relies on the resilience of the messaging platforms against impersonation attacks and session hijacking. 

Mallory can still send messages on behalf of Alice. This is not unrealistic as we assumed Mallory has access to the social network accounts of the survivor. Alice and Carol communicate their normal conversations through the messaging app and only use the encrypted messages in case of emergency. This is an agreement to minimise Mallory's suspicion.
Assuming Mallory is unsuccessful in guessing the key, he cannot decrypt the hidden message in the photo. Moreover, as cannot distinguish a regular photo from one with a hidden message based only on visual inspection, the adversary's options are limited.

\section{Conclusion}
\label{sec:concl}
This paper proposes three applications of proximity-based shared secrets supporting socially constrained communication. The proposed proximity-based shared secret generation method is a novel and secure solution that uses gyroscope data from the synchronized movement of smartphones as the basis for establishing a shared secret between two smartphones, without exchanging any data over the network. The evaluation results of the method show strong cryptographic properties, with the ability to generate nearly 8 bits of entropy per generated byte of data, which is similar to methods used within OpenSSL. The final secret generated after about 1 minute of shared movements on both phones has a length of around 300 bits, which can serve as the basis for a strong secret that can withstand reasonable attacks. As proof-of-concept, SecureWhispers App uses the generated secret to encrypt messages with standard message encryption techniques to hide users' messages from adversaries. In principle, the generated shared secret can also be used as a Key Derivation Function key in the Double Ratchet algorithm~\cite{perrin2016,alwen2019double}, to setup a Signal-like E2EE channel. 

These features make SecureWhispers a powerful abstraction for private communications that can support establishment of reliable communication channels for sensitive topics which vulnerable people may feel difficult to talk about in public given the surrounding social or political climate. Encouragement from a national regulator and other researchers motivates us to further improve the application and make it more robust for wider usage. Of course, as with many other PETs, such abstractions could also be used for subversive purposes, and so requires careful consideration when deploying in the real world.

\bibliographystyle{ieeetr}
\bibliography{bibfile}

\begin{thebibliography}{10}

\bibitem{10.1145/3424302.3425909}
R.~P. Jover, ``Security analysis of sms as a second factor of authentication: The challenges of multifactor authentication based on sms, including cellular security deficiencies, ss7 exploits, and sim swapping,'' {\em Queue}, vol.~18, p.~37–60, Sept. 2020.

\bibitem{heartfield2018protection}
R.~Heartfield and G.~Loukas, ``Protection against semantic social engineering attacks,'' {\em Versatile cybersecurity}, pp.~99--140, 2018.

\bibitem{8939840}
Hyunki-Kim, Jinhyeok-Oh, Changuk-Jang, Okyeon-Yi, Juhong-Han, Hansaem-Wi, and Chanil-Park, ``Analysis of the noise source entropy used in openssl’s random number generation mechanism,'' in {\em 2019 International Conference on Information and Communication Technology Convergence (ICTC)}, pp.~59--62, 2019.

\bibitem{sec_gen_rsa}
R.~L. Rivest, A.~Shamir, and L.~Adleman, ``A method for obtaining digital signatures and public-key cryptosystems,'' {\em Commun. ACM}, vol.~21, p.~120–126, feb 1978.

\bibitem{ipv_tech_sol}
A.~Z. Kouzani, ``Technological innovations for tackling domestic violence,'' {\em IEEE Access}, vol.~11, pp.~91293--91311, 2023.

\bibitem{sec_gen_ephi}
D.~S. Bhatti and S.~Saleem, ``Ephemeral secrets: Multi-party secret key acquisition for secure ieee 802.11 mobile ad hoc communication,'' {\em IEEE Access}, vol.~8, pp.~24242--24257, 2020.

\bibitem{secret_oo_thinair}
I.~Safaka, C.~Fragouli, K.~J. Argyraki, and S.~N. Diggavi, ``Creating shared secrets out of thin air,'' in {\em HotNets-XI}, 2012.

\bibitem{10.1145/3214275}
H.-M.~C. Leung, C.-W. Fu, and P.-A. Heng, ``Twistin: Tangible authentication of smart devices via motion co-analysis with a smartwatch,'' {\em Proc. ACM Interact. Mob. Wearable Ubiquitous Technol.}, vol.~2, July 2018.

\bibitem{10.1145/3264935}
S.~Mare, R.~Rawassizadeh, R.~Peterson, and D.~Kotz, ``Saw: Wristband-based authentication for desktop computers,'' {\em Proc. ACM Interact. Mob. Wearable Ubiquitous Technol.}, vol.~2, Sept. 2018.

\bibitem{10.1145/2684103.2684122}
R.~D. Findling, M.~Muaaz, D.~Hintze, and R.~Mayrhofer, ``Shakeunlock: Securely unlock mobile devices by shaking them together,'' in {\em Proceedings of the 12th International Conference on Advances in Mobile Computing and Multimedia}, MoMM '14, (New York, NY, USA), p.~165–174, Association for Computing Machinery, 2014.

\bibitem{10.1155/2021/6668478}
C.~Shi, L.~Xie, C.~Wang, P.~Yang, Y.~Song, S.~Lu, and Z.~Cai, ``Just shake them together: Imitation-resistant secure pairing of smart devices via shaking,'' {\em Wirel. Commun. Mob. Comput.}, vol.~2021, Apr. 2021.

\bibitem{9417366}
X.~Fang, L.~Liu, S.~Jia, X.~Zhao, and Y.~Zhang, ``P-shake: Towards secure authentication and communication between mobile devices,'' in {\em 2021 IEEE Wireless Communications and Networking Conference (WCNC)}, pp.~1--7, 2021.

\bibitem{10.1145/3378669}
Y.~Shen, B.~Du, W.~Xu, C.~Luo, B.~Wei, L.~Cui, and H.~Wen, ``Securing cyber-physical social interactions on wrist-worn devices,'' {\em ACM Trans. Sen. Netw.}, vol.~16, Apr. 2020.

\bibitem{mare:csaw19}
S.~Mare, R.~Rawassizadeh, R.~Peterson, and D.~Kotz, ``Continuous smartphone authentication using wristbands,'' in {\em Proceedings of the Workshop on Usable Security (USEC)}, Internet Society, February 2019.

\bibitem{sec_gen_walkie_talkie}
W.~Xu, G.~Revadigar, C.~Luo, N.~Bergmann, and W.~Hu, ``Walkie-talkie: Motion-assisted automatic key generation for secure on-body device communication,'' in {\em 2016 15th ACM/IEEE International Conference on Information Processing in Sensor Networks (IPSN)}, pp.~1--12, 2016.

\bibitem{sec_gen_fuzy_logic}
G.~Revadigar, C.~Javali, W.~Xu, A.~V. Vasilakos, W.~Hu, and S.~Jha, ``Accelerometer and fuzzy vault-based secure group key generation and sharing protocol for smart wearables,'' {\em IEEE Transactions on Information Forensics and Security}, vol.~12, no.~10, pp.~2467--2482, 2017.

\bibitem{sec_gen_shake2com}
Q.~Jiang, X.~Huang, N.~Zhang, K.~Zhang, X.~Ma, and J.~Ma, ``Shake to communicate: Secure handshake acceleration-based pairing mechanism for wrist worn devices,'' {\em IEEE Internet of Things Journal}, vol.~6, no.~3, pp.~5618--5630, 2019.

\bibitem{sec_gen_sens_pair}
J.~Han, A.~J. Chung, M.~K. Sinha, M.~Harishankar, S.~Pan, H.~Y. Noh, P.~Zhang, and P.~Tague, ``Do you feel what i hear? enabling autonomous iot device pairing using different sensor types,'' in {\em 2018 IEEE Symposium on Security and Privacy (SP)}, pp.~836--852, 2018.

\bibitem{10.1145/3023954}
W.~Xu, C.~Javali, G.~Revadigar, C.~Luo, N.~Bergmann, and W.~Hu, ``Gait-key: A gait-based shared secret key generation protocol for wearable devices,'' {\em ACM Trans. Sen. Netw.}, vol.~13, Jan. 2017.

\bibitem{10211219}
A.~Vyas, J.~Kim, and Q.~Wu, ``Localization information based secret key generation in chain and hybrid topology based wireless mobile network to secure multimedia data,'' in {\em 2023 IEEE International Symposium on Broadband Multimedia Systems and Broadcasting (BMSB)}, pp.~1--5, 2023.

\bibitem{9442833}
X.~Hu, L.~Jin, K.~Huang, X.~Sun, Y.~Zhou, and J.~Qu, ``Intelligent reflecting surface-assisted secret key generation with discrete phase shifts in static environment,'' {\em IEEE Wireless Communications Letters}, vol.~10, no.~9, pp.~1867--1870, 2021.

\bibitem{10008695}
E.~O. Torshizi and W.~Henkel, ``Reciprocity and secret key generation for fdd systems using non-linear quantization,'' in {\em 2022 IEEE Globecom Workshops (GC Wkshps)}, pp.~927--932, 2022.

\bibitem{10.1007/978-3-540-74853-3_18}
D.~Bichler, G.~Stromberg, M.~Huemer, and M.~L{\"o}w, ``Key generation based on acceleration data of shaking processes,'' in {\em UbiComp 2007: Ubiquitous Computing} (J.~Krumm, G.~D. Abowd, A.~Seneviratne, and T.~Strang, eds.), (Berlin, Heidelberg), pp.~304--317, Springer Berlin Heidelberg, 2007.

\bibitem{5227986}
D.~Bichler, G.~Stromberg, and M.~Huemer, ``Synchronizing shaking sequences for generating symmetric keys,'' in {\em 2009 2nd International Workshop on Nonlinear Dynamics and Synchronization}, pp.~75--80, 2009.

\bibitem{nist_800_131A}
E.~B. (NIST) and A.~R. (NIST), ``{Transitioning the Use of Cryptographic Algorithms and Key Lengths}.'' \url{https://csrc.nist.gov/pubs/sp/800/131/a/r2/final}, 2008.
\newblock [https://doi.org/10.6028/NIST.SP.800-131Ar2].

\bibitem{tseng20ipsinfidelityforums}
E.~Tseng, R.~Bellini, N.~McDonald, M.~Danos, R.~Greenstadt, D.~McCoy, N.~Dell, and T.~Ristenpart, ``The tools and tactics used in intimate partner surveillance: An analysis of online infidelity forums,'' in {\em 29th USENIX Security Symposium (USENIX Security 20)}, pp.~1893--1909, USENIX Association, Aug. 2020.

\bibitem{ceccio23spydevices}
R.~Ceccio, S.~Stephenson, V.~Chadha, D.~Y. Huang, and R.~Chatterjee, ``Sneaky spy devices and defective detectors: The ecosystem of intimate partner surveillance with covert devices,'' in {\em 32nd USENIX Security Symposium (USENIX Security 23)}, (Anaheim, CA), pp.~123--140, USENIX Association, Aug. 2023.

\bibitem{stephenson23injail}
S.~Stephenson, M.~Almansoori, P.~Emami-Naeini, and R.~Chatterjee, ``"{I}t{\textquoteright}s the equivalent of feeling like you{\textquoteright}re in {Jail{\textquotedblright}}: Lessons from firsthand and secondhand accounts of {IoT-Enabled} intimate partner abuse,'' in {\em 32nd USENIX Security Symposium (USENIX Security 23)}, (Anaheim, CA), pp.~105--122, USENIX Association, Aug. 2023.

\bibitem{sensor_rand}
M.~Hillerstr{\"o}m, I.~Ullah, and P.~J.~M. Havinga, ``Sensor-based puf: A lightweight random number generator for resource constrained iot devices,'' in {\em Internet of Things. IoT through a Multi-disciplinary Perspective} (L.~M. Camarinha-Matos, L.~Ribeiro, and L.~Strous, eds.), (Cham), pp.~89--105, Springer International Publishing, 2022.

\bibitem{9458231}
H.~Carlsson, I.~Skog, T.~B. Schön, and J.~Jaldén, ``Quantifying the uncertainty of the relative geometry in inertial sensors arrays,'' {\em IEEE Sensors Journal}, vol.~21, no.~17, pp.~19362--19373, 2021.

\bibitem{10227897}
G.~He, M.~Liu, X.~Li, H.~Chang, W.~Yuan, and G.~Yuan, ``Improving the position accuracy of the zupt-aided pedestrian inertial navigation by using a differential layout mimu array,'' {\em IEEE Sensors Journal}, vol.~23, no.~19, pp.~23420--23430, 2023.

\bibitem{sensor_drift}
F.~Wang, L.~Ling, and Z.~Qi, ``Research on data drift suppression method of gyroscope sensor based on first-order inertia and low pass filter,'' in {\em 2020 Chinese Control And Decision Conference (CCDC)}, pp.~1237--1241, 2020.

\bibitem{7512040}
R.~Jan, B.~Peter, and M.~Juraj, ``Comparison of output data from inertial sensors in smartphones,'' in {\em 2016 ELEKTRO}, pp.~83--87, 2016.

\bibitem{10147912}
G.~Patrizi, M.~Carratù, L.~Ciani, P.~Sommella, M.~Catelani, and A.~Pietrosanto, ``Analysis of inertial measurement units performances under dynamic conditions,'' {\em IEEE Transactions on Instrumentation and Measurement}, vol.~72, pp.~1--13, 2023.

\bibitem{7839934}
M.~Gong, T.~Zhan, P.~Zhang, and Q.~Miao, ``Superpixel-based difference representation learning for change detection in multispectral remote sensing images,'' {\em IEEE Transactions on Geoscience and Remote Sensing}, vol.~55, no.~5, pp.~2658--2673, 2017.

\bibitem{1524863fftspectralanalysis}
B.~Rust and D.~Donnelly, ``The fast fourier transform for experimentalists, part iv: autoregressive spectral analysis,'' {\em Computing in Science \& Engineering}, vol.~7, no.~6, pp.~85--90, 2005.

\bibitem{4762004}
D.~T. Lai, R.~Begg, E.~Charry, and M.~Palaniswami, ``Frequency analysis of inertial sensor data for measuring toe clearance,'' in {\em 2008 International Conference on Intelligent Sensors, Sensor Networks and Information Processing}, pp.~303--308, 2008.

\bibitem{10.1145/3442520.3442528}
N.~Lv, T.~Chen, and Y.~Ma, ``Analysis on entropy sources based on smartphone sensors,'' in {\em Proceedings of the 2020 10th International Conference on Communication and Network Security}, ICCNS '20, (New York, NY, USA), p.~21–31, Association for Computing Machinery, 2021.

\bibitem{perrin2016}
M.~M. Trevor Perrin~(editor), ``The double ratchet algorithm,'' 2016.
\newblock \url{https://signal.org/docs/specifications/doubleratchet/}.

\bibitem{alwen2019double}
J.~Alwen, S.~Coretti, and Y.~Dodis, ``The double ratchet: security notions, proofs, and modularization for the signal protocol,'' in {\em Annual International Conference on the Theory and Applications of Cryptographic Techniques}, pp.~129--158, Springer, 2019.

\end{thebibliography}

\appendix
\section{SecureWhispers App}
\label{sec:appdx_swa}
\begin{figure}[ht]
    \centering
    \includegraphics[width=0.45\linewidth]{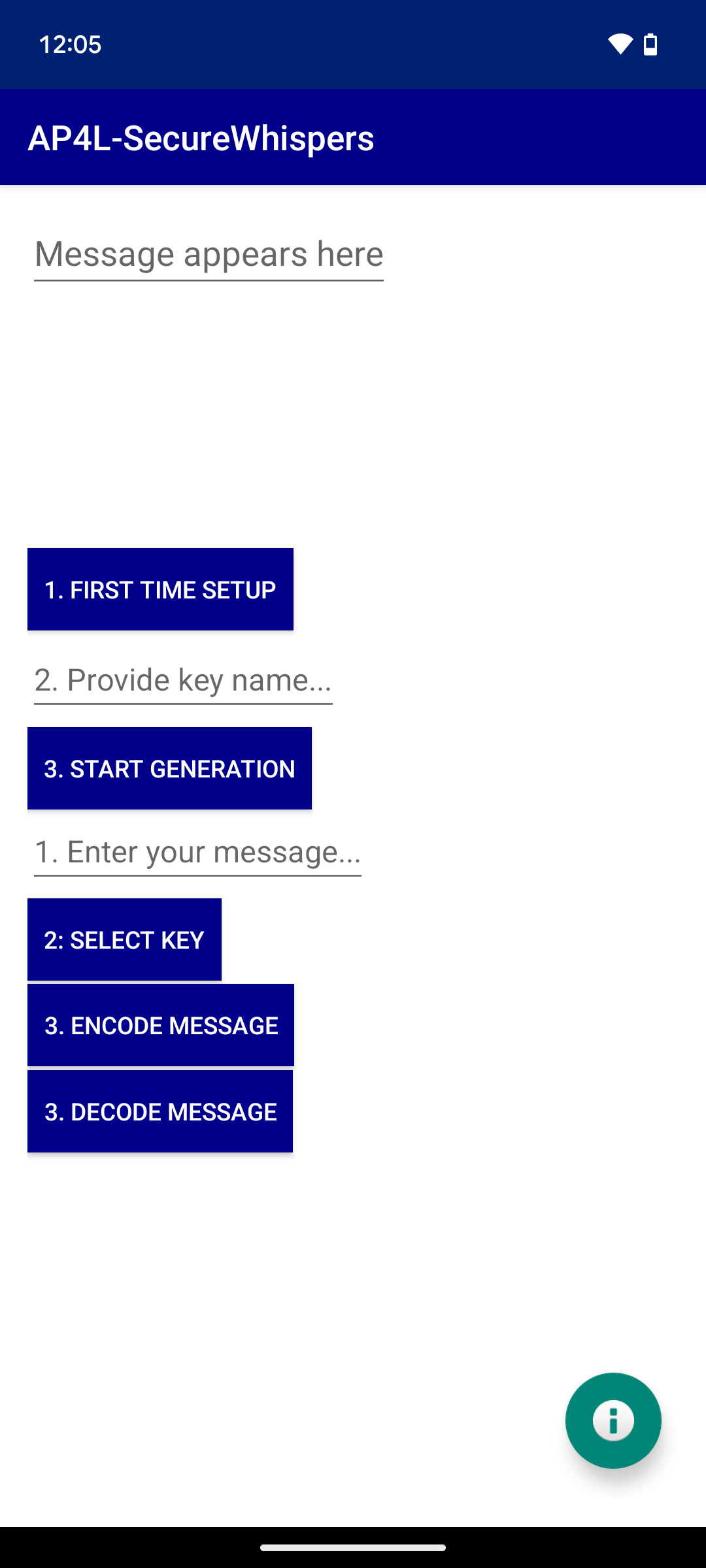}
    \caption{Screenshot of the ``SecureWhispers'' App}
    \label{fig:secwhis-scsh}
\end{figure}
The SsecureWhispers application implements our shared secret generation algorithm, user messages encryption using this secret, embedding the encrypted message into an image, and storing it on the device. Thus allowing the user to share encrypted message-embedded images through an external messaging application such as WhatsApp, Signal, or Telegram. A screenshot of the application is shown in Fig.~\ref{fig:secwhis-scsh}. Two trusted persons first generate a shared secret by means of synchronous movements of both smartphones. The sender side App uses this secret to secure user's data using the latest secret. The receiver side App then uses the shared secret to extract the secret message.

\end{document}